# Mixed Halide Perovskite Light Emitting Solar Cell

Dmitry Gets[1,*], Arthur Ishteev[1], Eduard Danilovskiy[1], Danila Saranin[3], Ross Haroldson[2], Sergey Makarov[1] and Anvar Zakhidov[1,2,*]

[1] *ITMO University, Department of Nanophotonics and Metamaterials, 191002, Russia, Saint Petersburg, Lomonosov str 9*
[2] *Alan G. MacDiarmid NanoTech Institute, University of Texas at Dallas, Richardson, TX 75083, USA*
[3] *National University of Science and Technology, NUST-MISIS, 119049 Russia, Moscow, Leninsky pr. 4*

*email: dmitry.gets@metalab.ifmo.ru
*email: zakhidov@utdalas.edu

# Abstract

Organic-inorganic halide perovskites recently have emerged as a promising material for highly effective light-emitting diodes (LEDs) and solar cells (SCs). Despite efficiencies of both perovskite SCs and LEDs are already among the best, the development of a perovskite dual functional device that is capable of working in these two regimes with high efficiencies is still challenging. Here we demonstrate that the dual functional device based on mixed halide perovskite $CH_3NH_3PbBr_2I$ can be switched from SC to LED with low threshold voltage $V_{th} < 2$ V by exposing to Sun at open circuit Voc or at small bias voltage of $V_{pol} \sim 1 - 2$ V. Such photo-poling creates in-situ p-i-n junction via methylammonium ($CH_3NH_3^+$, $MA^+$) and $I^-/Br^-$ ions migration to interfaces, lowering charge injection barriers, and self-balancing injection currents in perovskite LED. We show that before the photo-poling, the electroluminescence (EL) is highly unstable in LED regime, whereas after the photo-poling, stabilized EL exhibits unusual dynamics, increasing with time and poling cycle number, while $V_{th}$ and injection current decrease with cycling runs. Additionally, photo-induced and current-induced halide segregation accumulates with cycling, that is found beneficial for LED, increasing its efficiency and brightness, but reversibly degrading photovoltaic (PV) performance, which can be easily recovered.

## 1. Introduction

The organohalide perovskite solar cells (OHPS-SC) have shown superior efficiencies of > 23% which keeps growing [1], while at the same time perovskite LEDs keep demonstrating very high quantum efficiencies and bright pure colors [2]. This raised the question of creating the dual function device, which is a light emitting solar cell (LESC) that may show both good photovoltaic (PV) operation and good electroluminescence (EL) with high external quantum efficiency (EQEEL) in the same device. However, creation of such device is challenging, because achieving both good charge collection (PV operation) and efficient charge injection (LED operation) requires complex device design or use of properly doped transport layers with specific properties.

Several approaches were already demonstrated in order to achieve such dual functionality in one monolithic single halide perovskite device [3-6]. For example, the special type of ionic electron transport layer (ETL), such as PEI was suggested, which has ionic conductor properties and improves charge injection by doping OHPS/ETL interface [6]. In another example, authors utilized low work function Ba cathode [5]. But these dual functional devices either have low PCE or irradiate in IR region (while visible light EL is usually desired), or need complex device design to achieve high performance. Summary of earlier work on LESC is shown at Table 1 below with comparison to present approach.

Recently, it was demonstrated by Deng [7] that p-i-n structure could be formed inside perovskite layer by photo-poling. It was found that, device under 1 Sun illumination at photogenerated open circuit voltage ($V_{OC}$) creates an internal p-i-n junction in perovskite (PS) layer by moving ions of $MA^+$ and $I^-/Br^-$, towards PS/HTL and PS/ETL interfaces. As a result, $V_{OC}$ increases in 1 minute timescale and improves power conversion efficiency (PCE) of PV. This p-i-n structure quickly disappears by backward ionic migration, if light is switched off [7]. As we show below, similar in-situ p-i-n structure formation in mixed halide perovskite structures can be used not only for enhancement of PV characteristics, but also for LED operation improvement. Moreover, theoretical investigations [8] demonstrated importance of the optimization of transport layer parameters, particularly by adjusting the doping level and Fermi level position. Manipulation of these parameters [9, 10] are the ways for further performance increase of perovskite based devices.

Here we propose a different (than in previous [3-6]) strategy for cheap and effective dual functional LESC based on mixed halide OHPS-SC by demonstrating that photo-poling induced p-i-n junction can be used for switching from conventional SC mode ($V < V_{OC}$) to LED operation ($V_{bias} > V_{th}$) in a classical planar PV-structure with simple organic ETL. Usually PCBM or $C_{60}$ commonly used as ETL in SC, is not suitable for EL mode, owing to high barrier for electrons injection to conduction band of perovskite from low-lying lowest unoccupied molecular orbital (LUMO) of fullerides. In this study, to avoid this problem, we use mixed halide perovskite, since it has large enough band gap for EL in visible light range, and at the same time demonstrates enhanced ions migration. Although mixed halide perovskite exhibits the halide segregation effect under light soaking [11-14] or biasing [15], which usually degrades characteristics of PS, we show here benefits of segregation for LED operation. Indeed such mixed halide LESC exhibits brighter EL intensity growing with time and poling cycles due to anion segregation, since EL is governed by I-rich regions that collect radiative excitons by lower band gap [11]. This new photo-poling approach introduces the use of the internal ionic migration doping stages for tuning for better EL, similar to ionic diffusion in light emitting electrochemical cells (LEEC) [15, 16].

## 2. Experimental details

*2.1 Device Fabrication*

Perovskite based planar p-i-n devices were fabricated in nitrogen atmosphere. For the device fabrication was chosen the following scheme of functional layers ITO/PEDOT:PSS/Perovskite/PCBM/LiF/Ag. Device functional layers were subsequently deposited onto ITO covered glass substrates. Glass substrates with ITO pixels were cleaned in ultrasound bath in DI water, DMF, toluene, acetone and IPA consequently. Dried substrates were exposed to UV irradiation (189, 254nm) for 900s. Water dispersion of PEDOT:PSS 4083 was used as a hole transport material (HTL). It was filtered through PTFE 0.45 syringe filter and deposited by spin coating and annealed on a heating plate for 600s at 100°C in ambient atmosphere. Photoactive layer based on MAPbBr$_2$I was prepared by the consequent dissolving MAI (DyeSol) and PbBr$_2$ (Alfa Aesar "Puratronic" 99,999%) in DMF:DMSO (7:3) respectively. The solutions were steered overnight until dissolved at room temperature. Acquired perovskite inks were deposited by the single step solvent engineering technique [17] on top of HTL in 2 step spin-cycle. Diethyl ether was used as antisolvent and was slowly dripped on the rotating substrate in 10s past the jumping from 1000 to 2000rpm. Acquired perovskite films were stepwise annealed on the heating plate in dry atmosphere up to 100°C with 10°C step. Fullerene related material PC$_{60}$BM or C$_{60}$ were used as electron transport layer (ETL). PC60BM was dissolved in CB (20 mg/ml) and filtered through PTFE 0,45 syringe filter. Filtered solution was deposited onto perovskite layer by the spin coating. Layers of LiF (2nm) and the Ag cathode were deposited onto PC$_{60}$BM. C60 was thermally evaporated in the vacuum chamber. The thicknesses of the ETL, perovskite and HTL were measured by stylus profilometry. The thickness of Ag (C$_{60}$) cathode was in-situ controlled by Inficon deposition controller.

*2.2 Device characterization*

PV characterization was performed by measurements of J-V characteristics (Keithley 2400) of the solar cells. Devices were illuminated through apertures by solar simulator HAL-320 Asahi Spectra with an Air Mass 1.5 Global (AM 1.5G) spectra. LED measurements were performed using a Keithley 2400 source-measure unit, fiber spectrometer (Avantes AvaSpec mini) and Ophir Nova II optical power meter.

## 3. Results and discussion

The switching from PV to LED regime we first demonstrate by multiple cycling of J-V measurements runs of the PV structure based on a mixed halide MAPbBr$_2$I perovskite with a band gap of 2eV, aiming to achieve EL in visible range. These results are presented at Fig. 1 showing that stable and quite bright LED can be turned on after J-V runs in PV mode under 1 sun illumination.

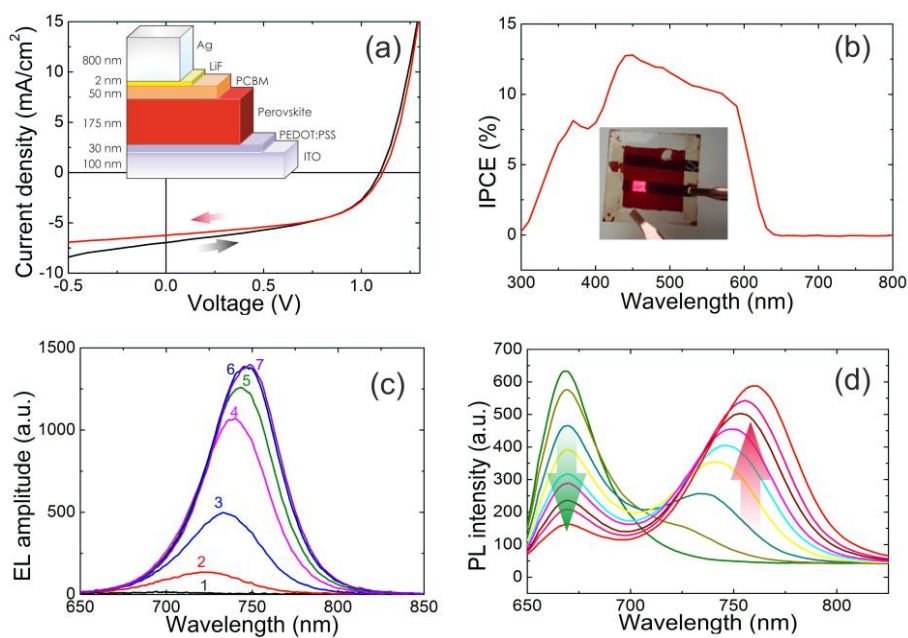

Fig. 1. (a) J-V curves in PV mode and structure of planar LESC device (b) incident photon to current efficiency (IPCE) spectral sensitivity of SC before poling (inset shows a photo of stable LED mode of the device after J-V cycling from -0.5 to + 1.5 V). (c) Electroluminescence spectra demonstrate time depended intensity enchancement after switching from SC to LED mode at $V_{th}$ = 2V. EL curves 1 – 7 have time step 5 seconds (d) 1-minute photoluminescence time dynamic, showing a strong segregation effect in mixed halide perovskite film.

The very first experiments performed at thin 175 nm PS film already demonstrated that after cycling in PV regime the device started to give quite bright and stable EL (inset on Fig. 1b). It was evident that exposure to light and some small voltage, can initiate EL in initial PV structure. Without exposure to light the EL regime was either not possible, or if pushed to light up at high threshold voltage ($V_{th}$) higher than 4 V the device quickly degraded in 1 minute (Fig. S1a). On the contrary, after photo-poling by cycling of J-V runs, the EL was stable and even showed increasing intensity in time. EL spectrum shifted from initial peak at 690 nm to red and was the brightest at 750 nm (Fig. 1c, S2 and S3), while IPCE of initial SC showed the large band gap Eg around 2eV (~600 nm) (Fig. 1b). The photoluminescence (PL) dynamics of mixed halide perovskite film clearly revealed the well-known segregation effect [11]: shift of the PL peak to the same wavelength range (~ 750 nm, Fig 1d and S12) caused by the domains enriched with I$^-$ ions and thus having lower band gap. Similarity of EL and PL dynamics clearly observed (Fig. 1c and d). The initial peak at 690 nm disappeared in EL that corresponds to fast and deep segregation enhanced by current injection [13].

First, we studied the evolution of PV parameters upon only photovoltage poling: no external voltage was applied or any J-V cycling done in PV. The real time increase of $V_{OC}$ was observed (inset in Fig. 2a). Once, J-V characteristic has been measured, the improvement was observed not only in $V_{OC}$, but also in fill factor (FF) and $J_{SC}$ (Fig.2a), that resulted in nearly twice improved PCE. This is well known effect of formation of p-i-n junction inside the PS layer upon poling first found by Deng [7, 14]. Dark J-V curves after $V_{OC}$ poling showed the step like log(J)-V curve, which reflected and confirmed the formation of p-i-n junction, with suppressed current at a small voltage below the internal barrier in p-i-n, and having a step when a voltage above p-i-n barrier (Figs. 2d and c). The increase of $V_{OC}$ upon p-i-n formation can be related to either better interface between n-doped PS and ETL by lowering the amount of interfacial traps, or by lowering the barrier at PS/ETL interface. Both these arguments imply the possible improved charge injection from the cathode. The photogenerated p-i-n is unstable in darkness and disappears, if PV is shorted [15]. After the relaxation, the new J-V demonstrated its initial low $V_{OC}$ = 0.6 V state and the dark J-V curve goes back to initial bad diode curve (black curve 1 at Fig. 2b) [7, 18].

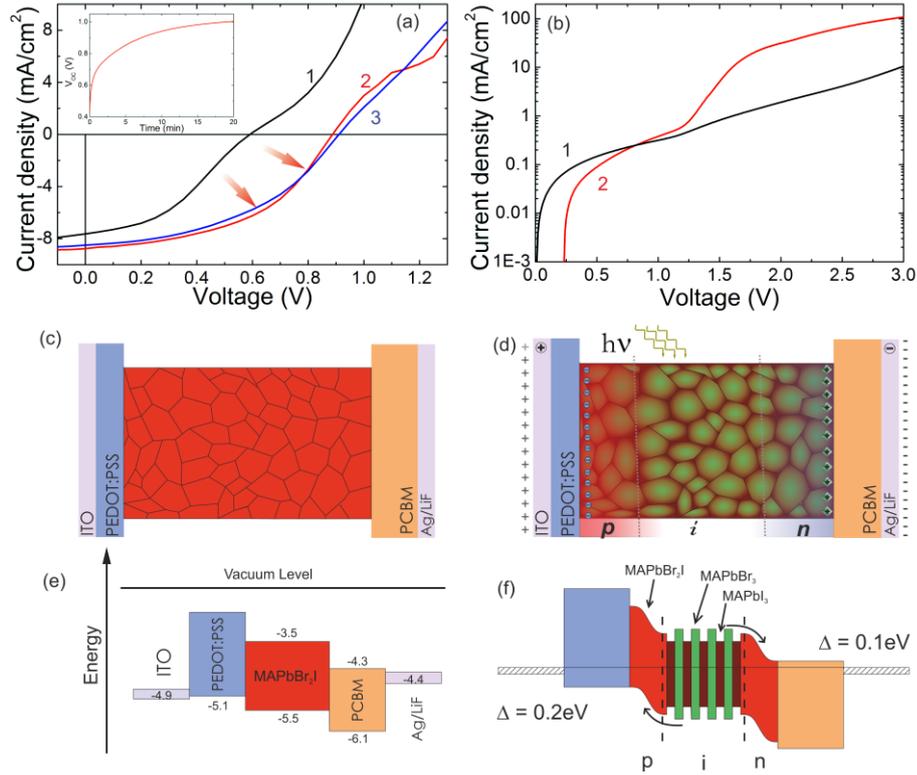

Fig. 2. (a) Self-improvement of $V_{OC}$ under light exposure in the open circuit regime. Inset shows the real time measurement of $V_{OC}$ increase under 1 sun. (b) The dark J-V curves of PV before (curve 1) and after self-poling (curve 2). P-i-n junction barrier typically has a "step" shape. Knee in log(J)-V curve corresponds to the straightening of p-n barrier. This allows charge injection in PS layer. (c) The original structure of PV without self-doping (Br- and I- ions distributed uniformly). (d) The p-i-n junction is formed by ionic diffusion to interfaces with transport layers. The i-area of p-i-n has segregated regions with I-rich ($E_g$ = 1.57 eV) and Br-rich ($E_g$ = 2.3 eV) nanograins, forming an internal bulk heterojunction. EL is enhanced from I-rich regions, by the exciton trapping and energy transfer from green (Br-rich) parts of bulk internal heterojunction. (e) Band diagram of non-segregated device. (f) Band diagram of segregated device: bulk heterojunction marked as alternating green-brown areas.

Next, for LED operation [18], we applied forward scan in small electric field and achieved switching to the LED mode by lighting the EL at small turn-on threshold voltage $V_{th}$ (Fig. 2b). The achieved stable EL after photo-poling demonstrated nontrivial time dynamics (Fig. 3 b-d). The combined effect of light enhanced ionic migration at higher poling bias of V = 1.5V and 2V and photoinduced segregation at prolonged exposure to sun illumination by cycling in PV regime was studied. We observed, that the poling at higher $V_{bias}$ aids to switch EL at lower $V_{th}$ (Fig. 3e), while $V_{th}$ is also lowered by larger number of cycle runs (Fig. 3e). We suggest that the in-situ n-doping of ETL by same ions: $MA^+$, $H^+$ and charged vacancies, of PS occurred at higher $V_{bias}$ [13, 20], and formed the external electronic doping of transport layers: with formation of extended p-(p-i-n)-n junctions. Similarly with I-, Br- ions and corresponding charged vacancies, penetrated into the PEDOT-PSS layer by ionic diffusion, induced by external high $V_{bias}$ > $E_g/e$ at sun, causing

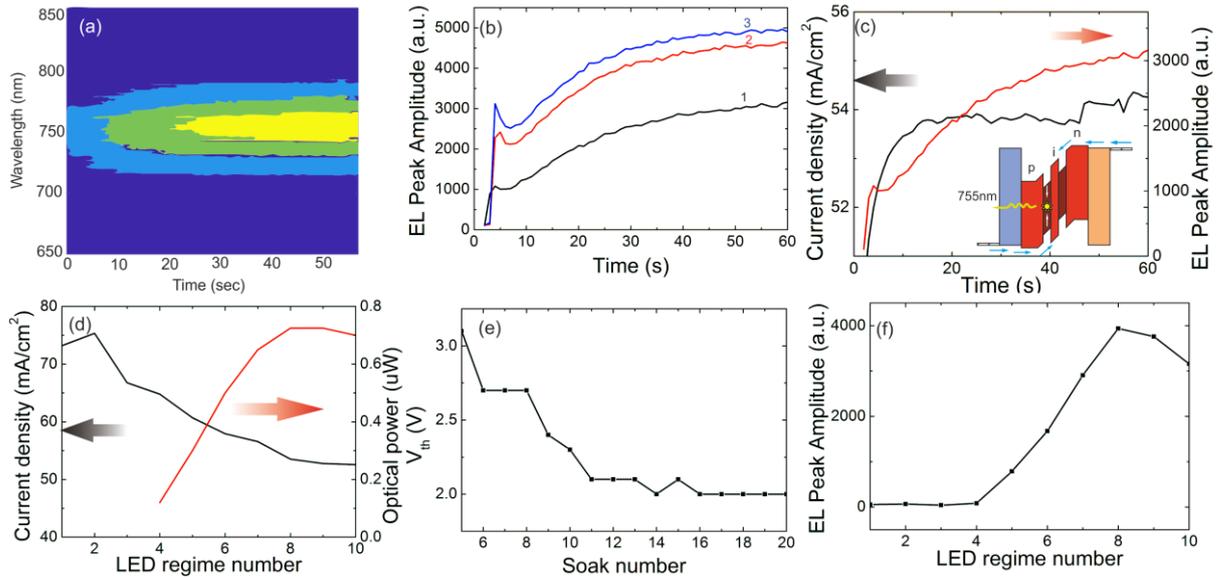

Fig. 3. LED regime turns on by multiple photo-poling cycles: (a) first LED lights up after several soaking cycles at +2V and 1 sun for 1 minute. (b) Enhanced EL amplitude in time measured after 3 soaks at +2V and 1 Sun. Signs indicate soak number. (c) Time depended current density and EL amplitude plot after soak #2 for LED regime. EL intensity was soared while current density saturated in LED regime after. This phenomena point to in-situ p-i-n structure development. (d) Current density and LED optical power evolve in the sequence of light soak - LED measurements at +2V. (e) The decrease of Vth value was observed upon 2V+ sun soak cycling. (f) Change in EL peak amplitude depends on number of "light soak – LED" cycles

the HTL p-type doping [20]. The electrochemical ionic doping requires charge injection from electrodes (Fig. 4c). On the contrary, photodoping at $V_{OC}$ injects photogenerated charge carriers (Fig. 4b).

Since LED regime needs higher voltage and higher current densities compared to PV regime, the first task was to keep photodoped p-i-n structure unrelaxed and load it with higher forward $V_{bias}$ in order to turn on the EL mode. Keeping this in mind, the following cycling was studied: after photo-poling at $V_{bias}$, the external voltage was applied to observe EL. Since p-i-n structure was stabilized by external $V_{bias}$ the forward scan could be performed until higher V of 4-5 Volts to observe the EL turning on at certain Vth. Indeed, EL was detected at low $V_{th}$ (Fig 3e) and this cycling of device regimes: photo-poling – LED scan – PV scan can be repeated many times (Fig. S4-S6).

Even a non pre-biased structure can give some unstable EL with very low intensity at high $V_{bias}$ > 3.5V or above, which then shows some orange/red light at a very high current of 50 mA. (Fig. S2). In this regime, to overcome the high initial barrier for electron injection at the very high $V_{th}$ ~ 4V creates a high unbalanced current, which is mostly hole current. Since device band structure now isn't optimized for charge injection the obtained luminosity and efficiency are very low. The boosted high hole current quickly overheats the device in 1-2 minutes and it completely degrades. Therefore, fast turned to LED mode PV device, without preliminarily pre-formed p-i-n junction makes the observed low EL absolutely unpractical, since device burns away very fast (Fig. S1a).

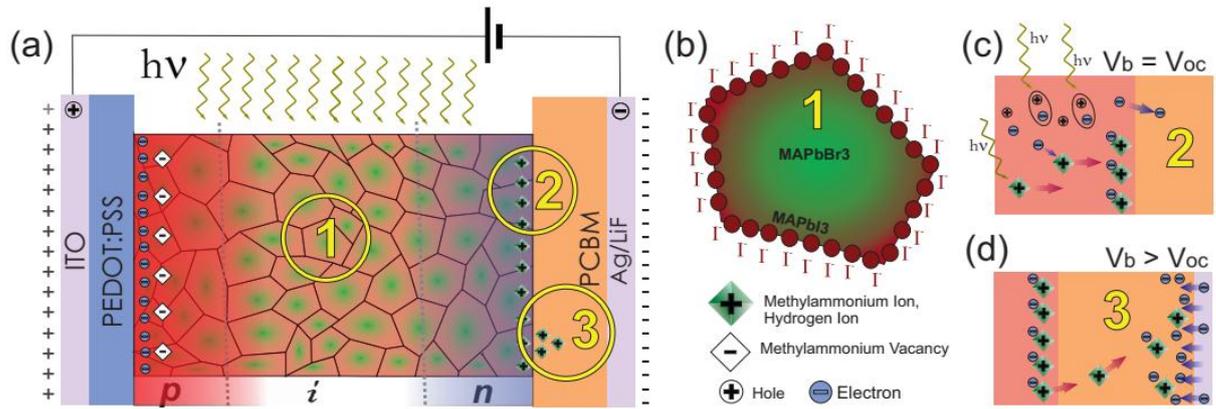

Fig. 4. The device segregation: (a) Schematic illustration of segregation process in the device under 1 sun illumination and applied external voltage. (b) Nanograins of the mixed halide perovskite film undergo segregation upon light and voltage soaking with the development of Br-rich and I-rich regions. (c) Methylammonium and hydrogen accumulation on the interface at V = $V_{OC}$. (d) Possible Methylammonium penetration into PCBM layer, causing n-doping of ETL at V > $V_{OC}$ via electrochemical charge injection, as assumed origin of improved EL operation at higher bias.

On the other hand, the device with the slowly prepared p-i-n structure by photo-poling showed stable EL with self-improving properties that are shown at Fig. 3c-d. The spectral shift in EL occurred in time rather fast: the initial EL peak at 690 nm decreases, while intensity of the peak at 750 nm corresponding to I-rich segregated regions increases by about 10 times.

The observed temporal dynamics of EL is quite unusual: EL intensity kept increasing in time during the first minutes of operation at certain $V_{op}$ = 2V. After every cycle of biasing under 1 sun illumination, the EL peak amplitude increased not only with time but also with cycle number (Fig. 3c). Moreover, the overall optical power also kept increasing from cycle to cycle, the $EQE_{EL}$ efficiency also increased to values of 0.024% (Fig. S7) which is comparable to $EQE_{EL}$ obtained in low work function Ba-cathode PV device made by Bolink [5].

The increased EL intensity can be explained by two factors: the segregation effect creates the I-rich areas at the interfaces of grains [21], while the Br-rich areas are in the middle parts of grains. The injected carriers are trapped (as radiative excitons) in low band gaps of the I-rich parts (Fig 3), and serve as radiative recombination centers for excitons providing effective EL [22, 23].

Table 1. Comparison of different approaches to Light emitting solar cells in previously reported work and in present paper

| Structure of the device | PV Parameters | EL spectrum and Parameters | Mechanism of charge injection/collection | Ref. |
|---|---|---|---|---|
| PS: MAPbBr$_2$I ITO/PEDOT:PSS/PS/PCBM/LiF/Al | $V_{OC}$ ~ 1.05V $V_{OC}$ dynamically improves from 0.6 to 1 V) upon light exposure due to p-i-n formation PCE = 5-8% $V_{OC}$ = 1.08 V $J_{SC}$ = 8mA/cm2 FF = 56% | LED Color changes during operation λ: 690 -> 750 nm $EQE_{EL}$ ≈ 0.03% and amplitude grows with exposure to light due to segregation of I-rich clusters. Opt. power ~ 200 µW and grows in time | Internal p-i-n formation in PS layer is induced by photo-poling at $V_{OC}$ and higher $V_{bias}$ with light soaking cycling for electrical field and light enhanced ionic migration in mixed halide. | This work |
| PS: MAPbBr$_3$ ITO/PEDOT:PSS/PS/ PEIBIm4/Ag | PCE = 1.02% $V_{OC}$ = 1.05V $J_{SC}$ = 3.12mA/cm2 FF = 31% | λ ≈ 520–560nm L ≈ 8000cd/m$^2$ $EQE_{EL}$ = 0.12 % | Lowering barrier at ETL induced by the n-doping or by dipoles at the interface induced by PEIBIm4 | Kim, et al. Ref.[6] |
| PS: MAPbI$_3$ FTO/TiO$_2$/m-TiO$_2$/PS/Spiro-OMETAD/Au | PCE = 20.8 % $V_{OC}$ = 1.16V $J_{SC}$ = 24.6mA/cm2 FF = 73% | λ = 775nm $EQE_{EL}$ = 0.5% | Initial p-i-n structure created by pre-doped p-HTL and n-ETL | Dongqin Bi et. al., Ref [3] |
| PS: MAPbI3 ITO/PEDOT:PSS/pTPD/PS/PCBM/Ba/Ag | PCE = 12.8 % $V_{OC}$ = 1.08V $J_{SC}$ = 18.5mA/cm$^2$ FF = 64% | λ = 765nm $EQE_{EL}$ = 0.04% Opt. power = 100µW | Low work function Ba cathode for increased electron injection in LED | L. Gil-Escrig et. al. Ref. [5] |
| PS: CsPbI$_3$ FTO/TiO$_2$/PS/ Spiro-OMETAD /MoO$_x$/Al | PCE = 10.77% $V_{OC}$ = 1.23V $J_{SC}$ = 13.4mA/cm$^2$ FF = 65% | λ ≈ 700nm | Quantum dot photoactive and emissive layer, which facilitate charge injection in LED mode | Abhishek et. al. Ref. [4] |

Figure 3 shows schematics of the segregation influence on EL in the p-i-n junction, which becomes a multiple bulk heterojunction (Fig. 3b).

After switching the device from the PV regime to EL, the observed dynamics of EL is quite complicated, and several effects probably contribute to it. At first stage, the in-situ formation of p-i-n in the PS layer is critically important. However, applying high $V_{bias}$ and the cycling under high $V_{bias}$ poling had different impacts. As can be seen from the inset on Fig. 4: Upon biasing by $V_{OC}$ generated upon light soaking the ions migration mainly occurred in PS layer, and photogenerated electrons neutralize the accumulated MA$^+$ ions in the n-type photodoped region of PS layer. It should be noted that at this biasing condition, the injection of electrons from Ag cathode to PS isn't favorable due to high barrier at the interface of Ag with PS, therefore the positive charge of MA+ ions could be compensated only by photogenerated electrons inside PS photoactive layer, (since injected electrons cannot support the n-doping within the PS layer). Therefore, the $V_{OC}$ biasing isn't effective to create full p-i-n junction and switch to LED mode, and at first stage, only the photogenerated electrons support the internal n-doping in one side of PS. Moreover, the light soaking enhances the migration of MA+ and other ions within PS layer, by the mechanism of lowering the activation barrier [20]. Upon small $V_{OC}$ bias, the photovoltage doesn't create strong enough internal electric E-field inside the ETL, and the ion fast migration within PS by photoenhanced diffusion has no driving force to penetrate further into PCBM layer. However, at higher $V_{bias}$ of 2V, the internal E-field in PCBM most probably pulls MA+ ions further into the PCBM layer. Also this is a hypothesis at this stage (not yet confirmed by direct measurement of ions present in ETL) for explaining EL improvement dynamics, we believe that this ETL doping is highly preferable by porous structure of fulleride ($C_{60}$ and PCBM) molecular solids This type of penetration of ions into PCBM [23] and into $C_{60}$ ETLs was demonstrated in [23-28] by external ionic gating in ionic liquids. The MA+ ions coming from PS have smaller size as compared to large ions of ionic liquids, and they can easily penetrate into PCBM, once there is a driving force for this. As demonstrated theoretically recently, the optimization of doping level and Fermi level position in transport layers crucially affects the performance of perovskite-based device [8]. Preliminary doping of the transport layers [9] or tuning of Fermi level position inside the transport layers [10] demonstrated good influence for device performance. Therefore, in-situ doping of transport layers can indeed aid for better dual functionality. As shown at Fig. 4, the ions are accumulated at the interface of ETL with Ag electrode and create another n-doped layer inside ETL by electrons injected from Ag electrode, since $V_{bias}$ = 2V is now higher than the electrochemical potential. This second stage doping of ETL (and similarly of HTL on other side) as compared to initial internal doping of PS layer itself, further lowers the interfacial barrier, decreases the series resistance, and is most likely to be a reason for the observed decrease of both Vth and injection current in LED regime.

    Thus, we can assume that the formation of secondary poling-induced doping of ETL and HTL creates (HTL p+) / (p-i-n) / (n+ ETL) structure with lower injection barriers. The migrated ions in ETL are trapped at defects, and stay in ETL longer time, then in parent PS, this is a most possible reason (which we are now trying to confirm by direct measurements by XRD and other methods) for increased EL intensity, optical power, and efficiency of EL after multiple poling cycling runs.

Table 1 represents the comparison of different approaches for development of dual functional LESC device.

## 4. Conclusion

The photo-poling of mixed halide PS-SC has been found to be a simple way to switch to LED regime of planar PV device by initiating ionic migration and in-situ doping by internal ions. The internal mobile ions within the PS layer redistribute under light soaking, creating both segregation and interfacial photodoped regions, and form p-i-n junction favorable for LED operation. Such LED is similar to light emitting electrochemical cells, with internal ions causing the photo-doping of PS layer upon $V_{OC}$ photovoltage. The secondary doping of transport layers (both ETL and HTL) at higher V poling and LED operating voltages, with further evolution into the $p^+$-(p-i-n)-$n^+$ structure is tentatively suggested to explain the observed dynamics of further EL improvement with time. Repeated multiple photo-poling cycles reversibly is decreasing solar cell performance and at the same time improving the brightness and $EQE_{EL}$ efficiency of LED.

To sum up, the proposed LESC device based on mixed halide perovskite can operate in a dual mode: first as a solar cell, with built in p-i-n junction upon light soaking, that improves its $V_{OC}$ and PCE; and then at higher forward $V_{pol} > V_{OC}$, it can be switched to the LED mode by poling. The removal of external bias leads to backward redistribution of ions to equilibrium $MAPbBr_2I$ in the bulk of mixed halide perovskite layer. Thus, the device can again operate in any of two different regimes. Since the PV recovers, its operation under overnight it is ready to operate again as SC, while further PV cycling upon light soaking helps to form a better LED operation. This study gives insight for development of novel energy saving devices like standalone lightning systems for signage or similar niche applications.

### Acknowledgment
The work was supported by the Ministry of Education and Science of the Russian Federation (Projects 16.8939.2017/8.9 and 2.2267.2017/4.6), Russian Foundation for Basic Researches (Project 18-33-00669) and partial support from the Welch Foundation (Grant AT 1617).